\documentclass[twocolumn,final,aps,prb,showpacs,superscriptaddress,amsmath,amssymb,amsfonts,floatfix]{revtex4-1}
\usepackage{graphicx}
\usepackage{multirow}
\usepackage{epsfig}

\newcommand{\ji}{\ensuremath{J_1}}
\newcommand{\tici}{\ensuremath{t_{\text{ic1}}}}
\newcommand{\jici}{\ensuremath{J_{\text{ic1}}}}
\newcommand{\ticii}{\ensuremath{t_{\text{ic2}}}}
\newcommand{\jicii}{\ensuremath{J_{\text{ic2}}}}

\newcommand{\tn}{\ensuremath{T_{\text{N}}}}
\newcommand{\mb}{\ensuremath{\mu_{\text{B}}}}
\newcommand{\mcr}{\ensuremath{m_{\text{Cr}}}}
\newcommand{\lix}{LiCr$X_2$O$_6$}
\newcommand{\lixx}{LiCr$X_2$O$_6$ ($X$\,=\,Si, Ge)}
\newcommand{\lisi}{LiCrSi$_2$O$_6$}
\newcommand{\lige}{LiCrGe$_2$O$_6$}
\def\be{\begin{equation}}
\def\ee{\end{equation}}
\newcommand{\mcc}[2]{\multicolumn{#1}{c}{#2}}
\usepackage{color}
\usepackage[colorlinks,breaklinks,bookmarks=true,citecolor=blue,linkcolor=red,urlcolor=blue]{hyperref}
\def\cred{\color{red}}
\def\cblue{\color{blue}}

\begin{document}

\title{Magnetic pyroxenes LiCrGe$_2$O$_6$ and LiCrSi$_2$O$_6$:\\ dimensionality
crossover in a non-frustrated $S$\,=\,$\frac{3}{2}$ Heisenberg model}

\author{O. Janson}
\email{janson@cpfs.mpg.de}
\affiliation{National Institute of Chemical Physics and Biophysics, 12618 Tallinn, Estonia}
\affiliation{Max-Planck-Institut f{\"{u}}r Chemische Physik fester Stoffe, N\"{o}thnitzer Str. 40, 01187 Dresden, Germany}

\author{G. N\'enert}
\affiliation{Institut Laue Langevin, Grenoble 38042, France}

\author{M. Isobe}
\affiliation{Institute for Solid State Physics, University of Tokyo, 5-1-5 Kashiwa, Chiba 277-8581, Japan}

\author{Y. Skourski}
\affiliation{Hochfeld-Magnetlabor Dresden, Helmholtz-Zentrum
Dresden-Rossendorf, D-01314 Dresden, Germany}

\author{Y. Ueda}
\affiliation{Institute for Solid State Physics, University of Tokyo, 5-1-5 Kashiwa, Chiba 277-8581, Japan}

\author{H. Rosner}
\affiliation{Max-Planck-Institut f{\"{u}}r Chemische Physik fester Stoffe, N\"{o}thnitzer Str. 40, 01187 Dresden, Germany}

\author{A.~A. Tsirlin}
\affiliation{National Institute of Chemical Physics and Biophysics, 12618 Tallinn, Estonia}
\affiliation{Max-Planck-Institut f{\"{u}}r Chemische Physik fester Stoffe, N\"{o}thnitzer Str. 40, 01187 Dresden, Germany}

\date{\today}

\begin{abstract}
The magnetism of magnetoelectric $S$\,=\,$\frac32$ pyroxenes LiCrSi$_2$O$_6$
and LiCrGe$_2$O$_6$ is studied by density functional theory (DFT) calculations,
quantum Monte Carlo (QMC) simulations, neutron diffraction, as well as
low-field and high-field magnetization measurements.  In contrast with earlier
reports, we find that the two compounds feature remarkably different, albeit
non-frustrated magnetic models.  In LiCrSi$_2$O$_6$, two relevant exchange
integrals, $J_1$\,$\simeq$\,9\,K along the structural chains and
$J_{\text{ic1}}$\,$\simeq$\,2\,K between the chains, form a 2D anisotropic
honeycomb lattice.  In contrast, the spin model of LiCrGe$_2$O$_6$ is
constituted of three different exchange couplings. Surprisingly, the leading
exchange $J_{\text{ic1}}$\,$\simeq$\,2.3\,K operates between the chains, while
$J_1$\,$\simeq$\,1.2\,K is about two times smaller.  The additional interlayer
coupling $J_{\text{ic2}}$\,$\simeq$\,$J_1$ renders this model 3D.  QMC
simulations reveal excellent agreement between our magnetic models and the
available experimental data.  Underlying mechanisms of the exchange couplings,
magnetostructural correlations, as well as implications for other pyroxene
systems are discussed.
\end{abstract}


\maketitle
\section{Introduction}

In the solid state, electricity and magnetism are related to two distinct
primary order parameters and pertain to different spontaneously broken
symmetries.  The interplay of the two orders is ruled by the magnetoelectric
(ME) coupling.  The underlying mechanism is still debated,\cite{[{See, e.g., }]
[{}] katsura2005,*chapon2006,*kimura2006,*choi2008}  yet a substantial progress
is achieved in material-specific studies.  Therefore, a viable way towards
better understanding is to pick a certain class of ME compounds realizing
different magnetic structures that would allow for a systematic study. 

Pyroxenes are one of the most promising candidates for such a systematic study.
This is a large group of natural minerals and inorganic compounds with a common
chemical formula $AMX_2$O$_6$, where the $A$ site can be occupied by alkaline
(Li and Na) or alkaline-earth (Mg and Ca) metals, the $M$ site can accommodate
various $3d$ transition metals, as well as Mg and Al. In natural pyroxenes, the
$X$ site is occupied by Si, while in most cases, the Ge counterpart
can be synthesized.

Several pyroxene materials were recently shown to exhibit ME properties.  For
instance, the natural mineral acmite NaFeSi$_2$O$_6$
(Ref.~\onlinecite{jodlauk2007}) and its Ge-containing counterpart
NaFeGe$_2$O$_6$ (Ref.~\onlinecite{kim2012}) are ME multiferroics.  ME effect
was also observed in LiFeSi$_2$O$_6$,\cite{jodlauk2007} as well as in
Cr-based pyroxenes LiCrSi$_2$O$_6$ (Ref.~\onlinecite{jodlauk2007})
and LiCrGe$_2$O$_6$ (Ref.~\onlinecite{nenert2010}).  To account for the ME
properties of these materials, and facilitate the search for new ME pyroxenes,
precise information on the magnetic model is essential. 

The magnetism of pyroxenes is ruled by the $M$ cations that form magnetic
chains of edge-sharing $M$O$_6$ octahedra.  Although these chains are common to
all pyroxenes, recent experiments reveal a variety of magnetic behaviors that
can substantially deviate from the quasi-one-dimensional (1D) chain physics.
For instance, the $S$\,=\,1/2 compound CaCuGe$_2$O$_6$ is a spin dimer
system.\cite{sasago1995} Another $S$\,=\,1/2 compound, LiTiSi$_2$O$_6$, also
has a singlet ground state (GS), but induced by orbital
ordering.\cite{isobe2002} On the other hand, the $S$\,=\,1 pyroxenes
LiVGe$_2$O$_6$ (Ref.~\onlinecite{[{}][{ and references therein.}]
blundell2003}) and LiVSi$_2$O$_6$ (Ref.~\onlinecite{[{}][{ and references
therein.}] pedrini2007}) exhibit long-range magnetic ordering, which is not
expected in isolated or weakly coupled $S=1$ chains showing Haldane physics.

The magnetic properties of pyroxenes are very sensitive to essentially
nonmagnetic constituents of the crystal structure. An instructive example is
given by $S$\,=\,3/2 pyroxenes with magnetic $M$\,=\,Cr$^{3+}$ atoms.  
The title compounds, \lisi\ and \lige, feature the same type of
antiferromagnetic order, albeit with different ordered magnetic
moments.\cite{nenert2009,nenert2010} The magnetic structure of NaCrSi$_2$O$_6$
is similar to \lisi\ and \lige,\cite{nenert2010} but the value of the ordered magnetic
moment (2.3\,\mb) in NaCrSi$_2$O$_6$ indicates its resemblance to the \lige\
germanate (2.33\,\mb), not the \lisi\ silicate (2.06\,\mb). However, its Ge
counterpart, NaCrGe$_2$O$_6$, is a ferromagnet with the ordered moment of only
$m_{\text{Cr}}$\,=\,1.85\,\mb, which is lower than in any other Cr$^{3+}$-based
pyroxene compound.\cite{nenert2009b}

The variety of magnetic behaviors observed in pyroxenes is rooted in their
electronic structure.  Using perturbation theory and density functional theory (DFT)
calculations, Streltsov and Khomskii studied the influence of the electronic
state of the magnetic $M$ cation (trivalent Ti, V, Cr, Mn, and Fe) and of the shortest
distance between the neighboring in-chain $M$ atoms on nearest-neighbor
magnetic exchange $J_1$.\cite{streltsov2008}  However, the intra-chain exchange
$J_1$ alone does not suffice to account for the magnetic GS, which is
influenced or even ruled by interchain couplings.  Based on geometrical
arguments, several authors alleged the presence of frustrated interchain
couplings,\cite{streltsov2008,nenert2010} yet no detailed investigation of the
microscopic magnetic model was performed so far. The main obstacles of this
task are the variety of possible superexchange paths and the low magnetic
energy scale (several K) of the ensuing magnetic couplings.  As a result, each
pyroxene compound requires a careful evaluation of its pertinent microscopic
parameters.

In this study, we perform a microscopic magnetic modeling for two ME pyroxenes,
\lisi\ and \lige.  They feature the same crystal and magnetic structure, yet
substantially different Curie-Weiss temperatures ($\theta$\,=\,34\,K and 6.5\,K in \lisi\
and \lige, respectively) and magnetic ordering temperatures (\tn\,=\,11.1\,K
and \tn\,=\,4.8\,K).  Moreover, \lisi\ shows a broad maximum in the magnetic
susceptibility at low temperatures, typical for the low-dimensional magnetism,
while the \lige\ closely resembles a three-dimensional antiferromagnet.
Finally, in the magnetically ordered GS, local magnetic moments are notably
different: 2.06(4) and 2.33(3)\,\mb/Cr in \lisi\ and \lige,
respectively.\cite{nenert2009,nenert2010}

To elucidate the origin of the different magnetic behaviors, we use a
combination of DFT calculations, quantum Monte Carlo (QMC) simulations, as well as
experimental neutron diffraction and high-field magnetization measurements. We
find that the magnetic properties of \lisi\ are described by an $S$\,=\,3/2
(isotropic) Heisenberg model on a 2D anisotropic honeycomb lattice.  The two
leading exchanges are \ji\,$\simeq$\,9\,K along the structural chains and
\jici\,$\simeq$\,2\,K operating via double bridges of SiO$_4$ tetrahedra
between the chains.  By contrast, \lige\ features, in addition to \ji\ and
\jici, another interchain coupling \jicii, also operating along the double
bridges of GeO$_4$ tetrahedra.  The
\ji\,:\,\jici\,:\,\jicii\,=\,0.5\,:\,1\,:\,0.5 ratio according to our
investigations indicates a 3D magnetism, with the leading interchain coupling
\jici\,$\simeq$\,2.3\,K.  Extensive QMC simulations for both microscopic models
reveal excellent agreement with all available experimental data, including the
hitherto never reported high-field magnetization curves.

This paper is organized as follows.  Sec.~\ref{S-method} contains information
on methodological and technical aspects of our DFT and QMC studies, as well as
neutron diffraction and magnetization measurements.  The crystal structure of
pyroxenes is briefly discussed in Sec.~\ref{S-str}. Microscopic DFT-based
modeling, QMC simulations of the resulting model and comparison with the
experiments are presented in Sec.~\ref{S-model}.  The ordered magnetic moment,
the mechanisms facilitating in-chain and interchain couplings, as well as
magnetostructural correlations are discussed in Sec.~\ref{S-disc}.  Finally,
the results are summarized in Sec.~\ref{S-summary}.

\section{\label{S-method}Methods}
Polycrystalline samples of \lisi\ and \lige\ were prepared by a solid-state
reaction with an appropriate molar ratio of Li$_2$CO$_3$, Cr$_2$O$_3$, and
GeO$_2$ (SiO$_2$). The weighted mixtures were pressed into pellets and heated
at 1273\,K in air for several days with one intermediate grinding.

\begin{figure}[tbp]
\includegraphics[width=8.6cm]{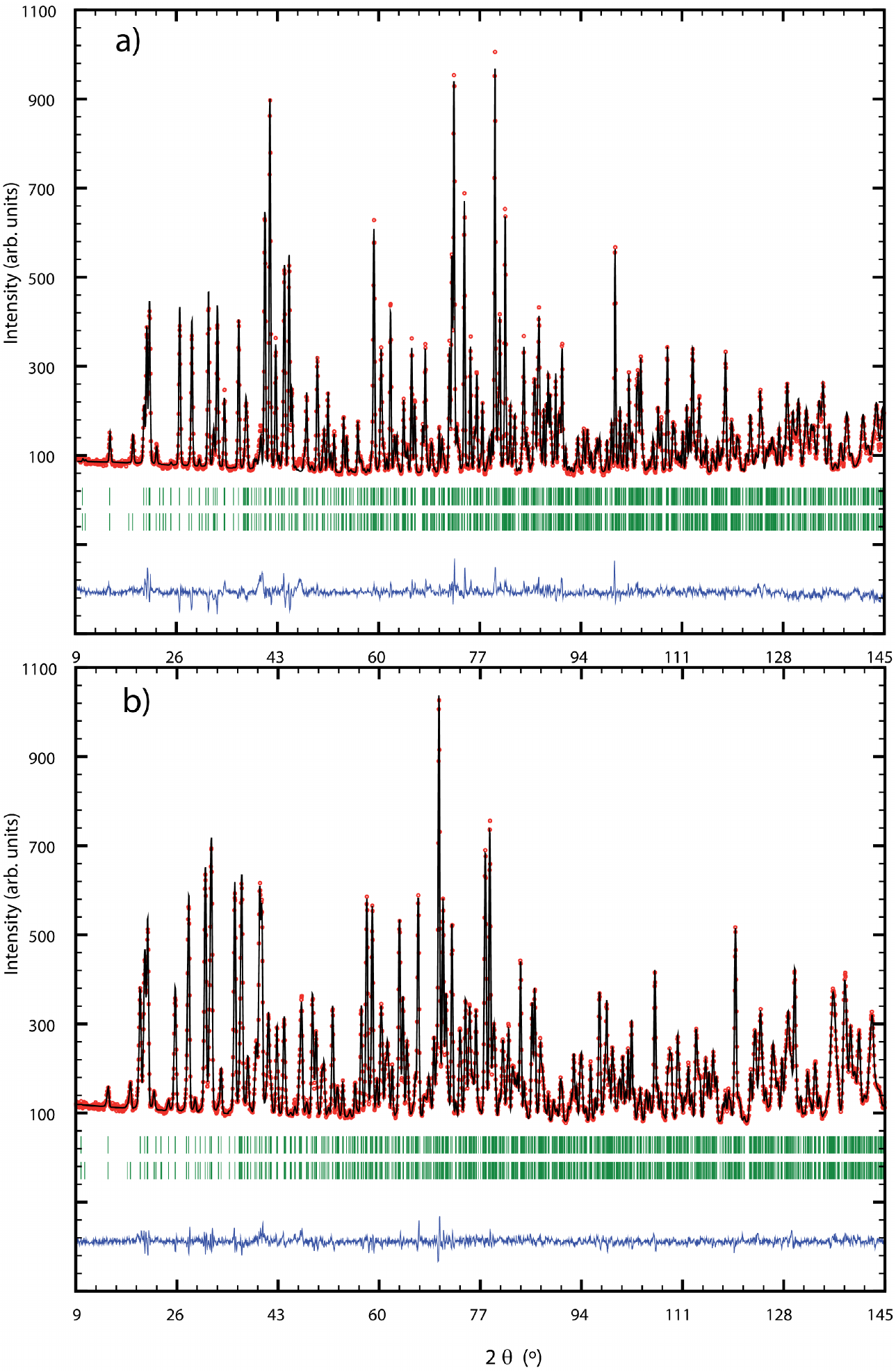}
\caption{\label{F-ND}(Color online) 
Neutron powder pattern ($\lambda$\,=\,1.594\,\r{A}) of \lisi\ (a) and \lige\
(b) collected at 1.8\,K using the D2B diffractometer.  The refinement has been
done in the $P2_1/c$ (14) space group with the following statistics:
$R_{\text{p}}$\,=\,4.55\% and $R_{\text{wp}}$\,=\,5.97\% for \lisi\ and
$R_{\text{p}}$\,=\,4.55\% and $R_{\text{wp}}$\,=\,1.97\% for \lige.}
\end{figure}

Neutron-diffraction measurements were carried out on powder samples. The
stoichiometry of the compounds as well as their precise crystal structures were
investigated using high resolution powder data at 1.8\,K using the D2B
diffractometer at the Institut Laue Langevin.  The measurements were carried
out at a wavelength of 1.594\,\r{A} corresponding to the (335) Bragg reflexion
of a germanium monochromator. The neutron detection is performed with $^3$He
counting tubes spaced at 1.25$^{\circ}$ intervals. A complete diffraction
pattern (Fig.~\ref{F-ND}) is obtained after about 25 steps of 0.05$^{\circ}$ in
2$\theta$.

Magnetic susceptibility was measured in a SQUID MPMS magnetometer in the
temperature range 1.8--380\,K in applied fields of 0.1\,--5\,T. High-field
magnetization curves were measured on powder samples at a constant
temperature of 1.5 K using a pulsed magnet at the Dresden High Magnetic Field
Laboratory (HLD), as described in Ref.~\onlinecite{tsirlin2009}.

DFT calculations were performed using two different codes: the full-potential
code \textsc{fplo9.03-37} (Ref.~\onlinecite{fplo}) 
, as well as the pseudopotential code \textsc{vasp-5.2}
(Ref.~\onlinecite{vasp_1, *vasp_2}).  For the exchange and correlation
potentials, we used both local density approximation\cite{pw92} (LDA) and
generalized gradient approximation\cite{pbe96} (GGA).  Strong electronic
correlations were treated in the DFT+$U$ scheme.  Alternatively, we used the
hybrid functionals PBE0 (Ref.~\onlinecite{pbe0}) and HSE06
(Ref.~\onlinecite{hse03, *hse06}) containing a fraction of the exact
(Hartree-Fock) exchange.  For \textsc{vasp-5.2} calculations, we used the
default projector-augmented wave (``PAW-PBE'') pseudopotentials.  For the
DFT+$U$ calculations, the fully localized limit (FLL) flavor of the
double-counting correction was used. The on-site repulsion and Hund's exchange
were fixed at $U_{3d}=3$\,eV and $J_{3d}$\,=\,1\,eV.\cite{janson2013} 

The experimental lattice constants and atomic coordinates were used
as a structural input.\cite{suppl} For non-magnetic calculations,
$\vec{k}$-meshes of 16$\times$16$\times$12 points (882 points in the
irreducible wedge) were used.  Wannier functions (WF) for the Cr $3d$ states
were evaluated using the procedure described in Ref.~\onlinecite{fplo_wf}.
Spin-polarized calculations were performed for magnetic supercells of two
types: (i) symmetry-reduced (space group $P1$) supercells, metrically
equivalent to the crystallographic unit cells (6$\times$6$\times$4
$\vec{k}$-points), and (ii) symmetry-reduced (space group $P1$) supercells
doubled along the $c$ axis (2$\times$2$\times$2 $\vec{k}$-points). 

QMC simulations were performed using the codes \textsc{loop}
(Ref.~\onlinecite{loop}) and \textsc{dirloop\_sse}
(Ref.~\onlinecite{dirloop_sse}) from the software package
\textsc{alps-2.1.1}.\cite{alps2.0, *alps1.3}  
All simulations were performed on finite lattices using periodic boundary
conditions.  Temperature dependencies of the magnetic susceptibility were
simulated on finite lattices of 1152 and 864 $S$\,=\,3/2 spins for \lisi\ and
\lige, respectively.  To simulate the magnetization process, finite lattices of
288 (\lisi) and 2048 (\lige) spins were used.  The magnetic ordering transition
temperature was estimated as the intersection of $N\rho_S(T)$ curves computed
for different finite lattices, where $\rho_S$ is the spin stiffness and $N$ is
the size of the finite lattice. The static structure factor $\mathbb{S}$ was
simulated on finite lattices of up 1024 spins.  

\section{\label{S-str}Crystal structure}

\begin{figure}[tbp]
\includegraphics[width=8.6cm]{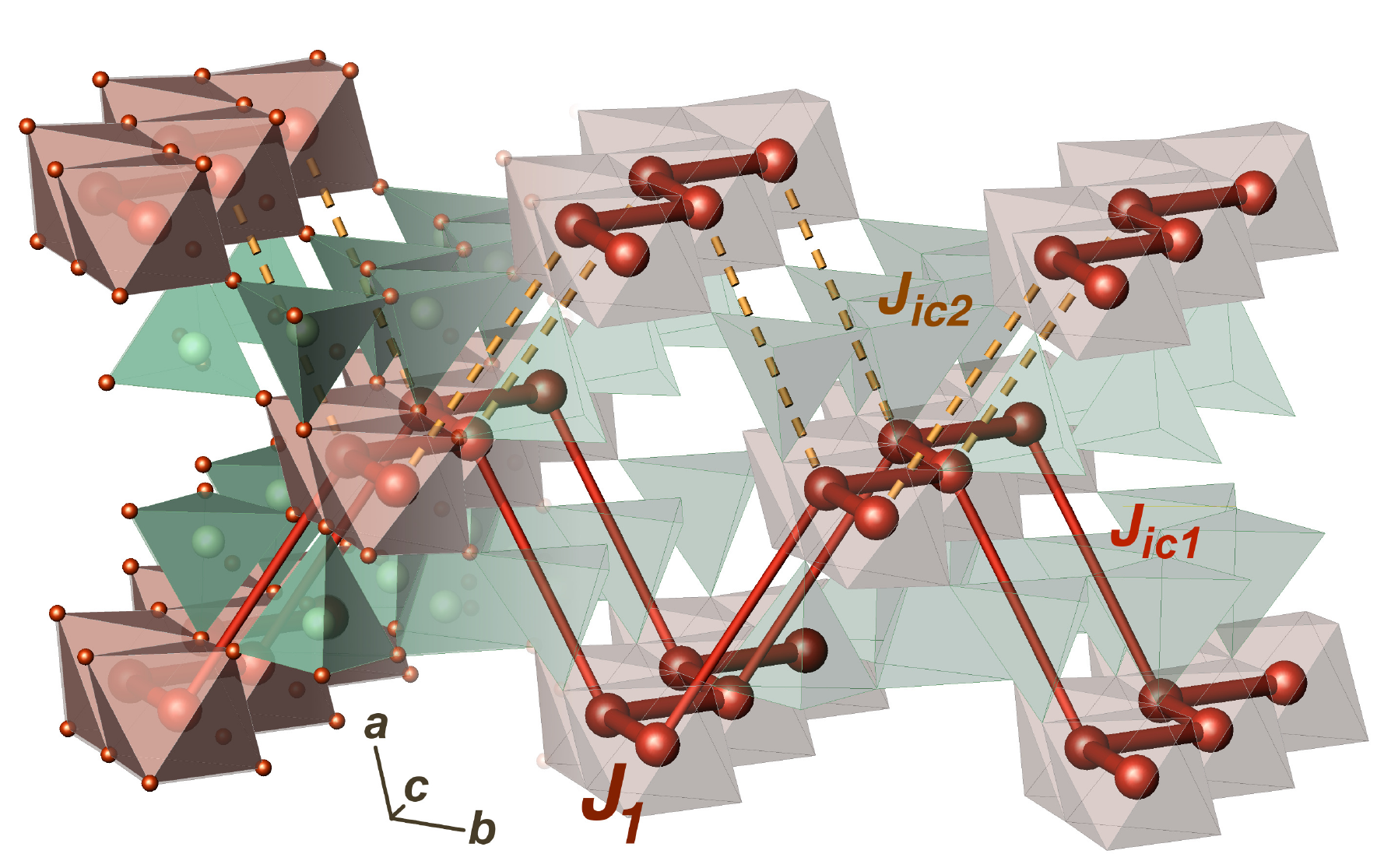}
\caption{\label{F-str}(Color online) Crystal structure and magnetic model of
\lixx. The crystal structure is formed by chains of edge-sharing CrO$_6$
octahedra (Cr atoms shown as large spheres) and chains of corner-sharing
SiO$_4$ tetrahedra running along $c$. The spin model features the
nearest-neighbor coupling \ji\ (thick solid lines) along the structural chains,
as well as two interchain couplings  \jici (thin solid lines) and \jicii\ (dashed
lines).  Note that \ji\ and \jici\ form a 2D anisotropic honeycomb lattice.}
\end{figure}

Two different monoclinic modifications of \lisi\ are known.  The
high-temperature phase crystallizes in the base-centered monoclinic space group
$C2/c$.\cite{redhammer2004} Below 330\,K, it transforms into the
low-temperature phase with the reduced symmetry
(sp.~gr.~$P2_1/c$);\cite{redhammer2004,nenert2009} the differences between the
two modifications are extensively discussed in Ref.~\onlinecite{redhammer2004}.
For \lige, the situation is more involved.  The low-temperature modification is
$P2_1/c$,\cite{nenert2010} but room-temperature measurements are controversial:
the authors of Ref.~\onlinecite{redhammer2008} refined their x-ray diffraction
data in the primitive space group, while Ref.~\onlinecite{matsushita2010}
reports the base-centered space group based on synchrotron measurements.
Investigation of the phase stability at room temperature is beyond the scope of
the present study, hence we restrict ourselves to the low-temperature $P2_1/c$
phases of \lisi\ and \lige.

The structure of the low-temperature modification is shown in Fig.~\ref{F-str}.
It is shaped by alteration of two chain-like elements: (i) cationic chains of
edge-sharing CrO$_6$ octahedra and (ii) anionic chains of corner-sharing XO$_4$
tetrahedra, both running along the crystallographic $c$ axis.  The symmetry
reduction toward $P2_1/c$ is essential for the magnetic properties: it gives
rise to inequivalent interchain couplings, in particular \jici\ and \jicii\
depicted in Fig.~\ref{F-str}.

As will be shown below, the magnetic properties of \lixx\ depend on subtle
structural details.  Therefore, for a microscopic DFT-based analysis, reliable
crystallographic information is a prerequisite.  Here, we use neutron powder
diffraction to determine the low-temperature crystal structures for both
compounds.  The resulting unit cell parameters and atomic coordinates are
provided in Table.~\ref{T-str}.

\begin{table}[tb]
\caption{\label{T-str} Atomic coordinates and isotropic displacement parameters
$U_{\text{iso}}$ (in 10$^{-2}$\,\r{A}$^{2}$) determined by neutron powder
diffraction ($\lambda$\,=\,1.594\,\r{A}) at 2\,K.  
The space group is $P2_1/c$\,(14).  The unit cell parameters are
$a$\,=\,9.7919(1)\,\r{A}, $b$\,=\,8.7149(1)\,\r{A}, $c$\,=\,5.33461(6)\,\r{A},
$\beta$\,=\,108.9146(6)$^{\circ}$ for \lige\ and $a$\,=\,9.5122(4)\,\r{A},
$b$\,=\,8.5713(4)\,\r{A}, $c$\,=\,5.2229(2)\,\r{A},
$\beta$\,=\,109.7569(7)$^{\circ}$ for \lisi.  All atoms occupy the $4e$
Wyckoff positions.}
\begin{ruledtabular}
\begin{tabular}{l l l l l}
 \mcc{5}{\lige}\\
Atom&  \mcc{1}{$x/a$} & \mcc{1}{$y/b$} &  \mcc{1}{$z/c$} & \mcc{1}{$U_{\text{iso}}$}\\
 Li &  0.2579(7)  &0.0138(6)  &0.2184(12)& 0.62(11)\\
 Cr &  0.2515(3)  &0.6593(4)  &0.2118(6) & 0.39(5)\\
 Ge1&  0.04756(14)&0.34430(17)&0.2750(3) & 0.38(2)\\
 Ge2&  0.55472(13)&0.84153(19)&0.2294(2) & 0.38(2)\\
 O1a&  0.85755(19)&0.3326(3)  &0.1740(3) & 0.50(2)\\
 O1b&  0.36416(19)&0.8316(3)  &0.1043(4) & 0.50(2)\\
 O2a&  0.11481(20)&0.5264(2)  &0.2840(4) & 0.50(2)\\
 O2b&  0.6312(2)  &0.0065(2)  &0.3874(4) & 0.50(2)\\
 O3a&  0.11786(19)&0.29075(20)&0.6099(4) & 0.50(2)\\
 O3b&  0.6136(2)  &0.6879(2)  &0.4544(4) & 0.50(2)\\ \hline
 \multicolumn{5}{c}{\lisi}\\
Atom&  \multicolumn{1}{c}{$x/a$} & \multicolumn{1}{c}{$y/b$} &
\multicolumn{1}{c}{$z/c$} & \multicolumn{1}{c}{$U_{\text{iso}}$}\\
 Li &  0.2517(10) &0.0120(7)  &0.2280(17)& 0.77(13)\\
 Cr &  0.2516(5)  &0.6579(5)  &0.2351(8) & 0.62(6)\\
 Si1&  0.0497(3)  &0.3412(4)  &0.2729(6) & 0.26(6)\\
 Si2&  0.5501(4)  &0.8409(4)  &0.2485(6) & 0.40(6)\\
 O1a&  0.8668(3)  &0.3325(3)  &0.1660(5) & 0.54(5)\\
 O1b&  0.3668(3)  &0.8355(3)  &0.1290(5) & 0.74(5)\\
 O2a&  0.1171(3)  &0.5116(3)  &0.3066(5) & 0.65(5)\\
 O2b&  0.6252(3)  &0.0056(3)  &0.3561(5) & 0.44(5)\\
 O3a&  0.1100(3)  &0.2692(3)  &0.5835(5) & 0.53(4)\\
 O3b&  0.6069(3)  &0.7171(3)  &0.5000(5) & 0.55(5)\\
\end{tabular}
\end{ruledtabular}
\end{table}

\section{\label{S-model}Microscopic magnetic modeling}

\subsection{DFT calculations}
\begin{figure}[tbp]
\includegraphics[width=8.6cm]{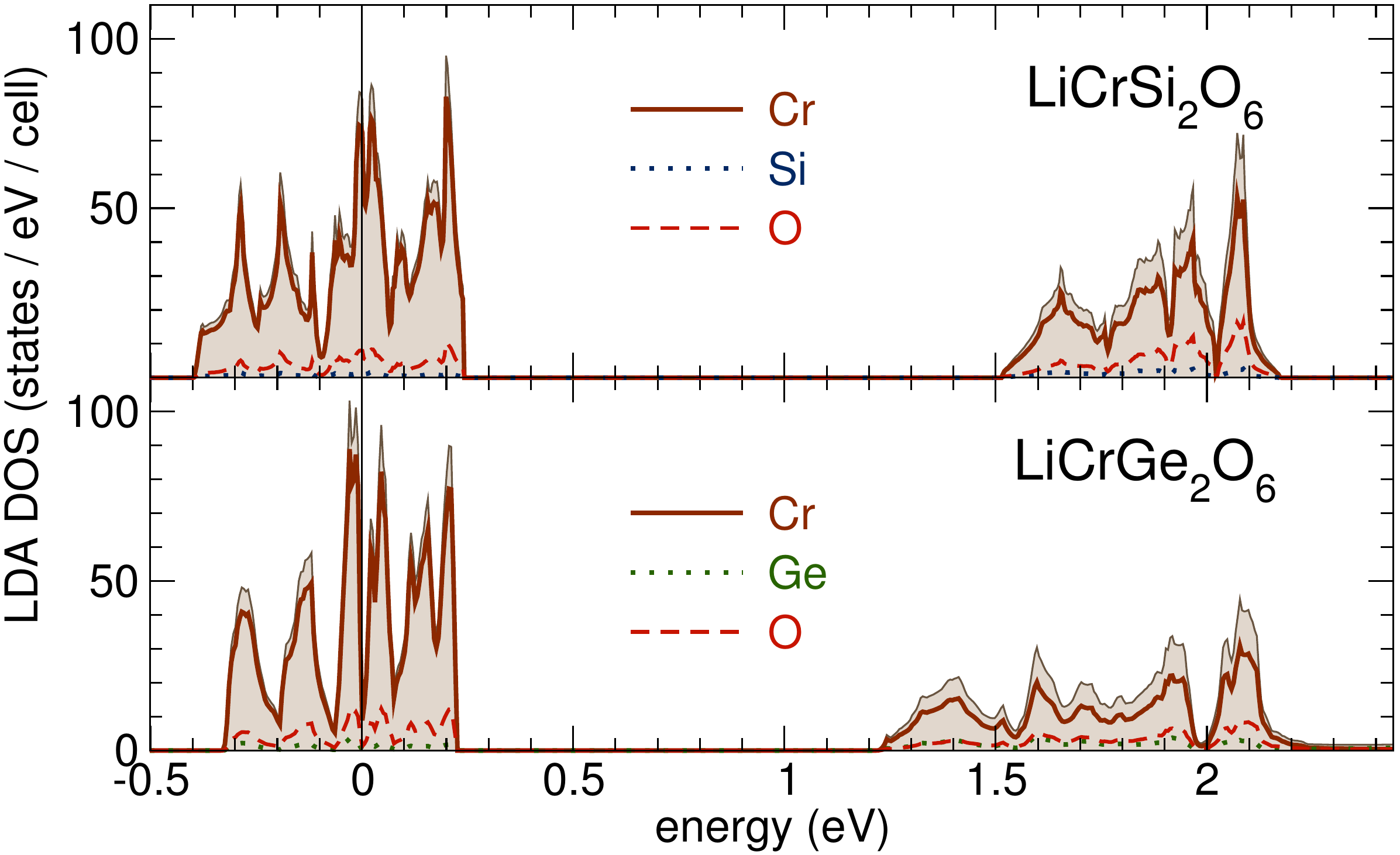}
\caption{\label{F-dos} (Color online) Nonmagnetic LDA total and atomic-resolved
density of states (DOS) for \lisi\ and \lige.  The Fermi level is at zero
energy.  GGA yields an almost indistinguishable DOS.
} \end{figure}

We start our analysis with nonmagnetic band structure calculations.  The
density of states (DOS) in both \lisi\ and \lige\ indicates the 3$d^3$
configuration of Cr$^{3+}$: the valence band comprises the half-filled $t_{2g}$
and the empty $e_g$ manifold, split by the crystal field (Fig.~\ref{F-dos}).
As expected for an octahedral coordination, the $e_g$ orbitals are mixed with O
$2p$ states, while the bands at the Fermi level are almost pure Cr $t_{2g}$
states.  The sizable exchange splitting, typical for Cr$^{3+}$,  splits the
$t_{2g}$ manifold into well-separated spin up and spin down densities and
ensures the high-spin configuration, readily obtained in spin-polarized
calculations.

The magnetism of Cr$^{3+}$ pyroxenes is driven by two concurrent processes: the
AF exchange ensuing from the hopping between the half-filled $t_{2g}$ orbitals,
and the FM exchange due to the hopping between half-filled $t_{2g}$ to the
empty $e_g$ orbitals.\cite{streltsov2008} 
To estimate the hopping integrals between different $d$ orbitals of Cr, we
construct Wannier functions and evaluate the on-site energies and transfer
integrals as diagonal and non-diagonal elements, respectively.  For both
compounds, we find only three paths that substantially contribute to the
electron transfer.  In accord with the previous DFT
studies,\cite{streltsov2008} the dominant coupling $t_1$ operates along the
structural chains.  Two other couplings, \tici\ and \ticii, are long-range and
operate via double bridges of GeO$_4$ or SiO$_4$ tetrahedra.  The full hopping
matrices for these couplings are provided in Supplementary
information.\cite{suppl}

Comparison of different hopping processes allows us to perform a qualitative
analysis and estimate the leading exchange couplings. We start with the
nearest-neighbor exchange \ji.  In \lisi, the strongest hoppings within the
$t_{2g}$ manifold occurs between the $yz$ and the $xz$ orbital of the
neighboring Cr atom (135\,meV).   The ensuing strong AF exchange is balanced
by a sizable hopping to the empty $x^2$\,$-$\,$y^2$ orbital (140\,meV), which
contributes to the FM exchange.  The situation in \lige\ is different: here,
the strongest hopping occurs between the $xz$ and the $e_g$ orbitals, while the
hoppings within the $t_{2g}$ manifold are suppressed ($<$100\,meV).  This
difference is responsible for the suppression of \ji\ in the Ge system.

The interchain coupling is realized along the two inequivalent paths, involving
double bridges of anionic tetrahedra. Each of the interchain couplings \tici\ and
\ticii\ is dominated by a single matrix element from the $t_{2g}$ manifold
amounting to 60--70\,meV, and two sizable hoppings between the $t_{2g}$ and
$e_g$ states.  The small energy scale of these concurrent processes impedes a
reliable estimation of the resulting exchange. For a more robust, although
still qualitative estimate, we turn to total energy DFT+$U$ and hybrid
functional calculations.

A well-known drawback of conventional DFT approaches (LDA and GGA) is a severe
underestimation of electronic correlations within the $3d$ shell of Cr$^{3+}$,
leading to underestimated band gaps\cite{mazin2007} and overestimated
exchange integrals.\cite{janson2013}  The simplest, yet widely renowned
approach to mend this problem is the DFT+$U$ method, which accounts for the
correlations by treating the on-site Coulomb repulsion $U_d$ and the on-site
Hund's exchange $J_d$ between the $d$ electrons in a mean-field fashion.  

Here, we adopt $U_d$\,=\,3.0\,eV and $J_d$\,=\,1.0\,eV that accurately
reproduce the experimental magnetic behavior for the recently studied quasi-1D compound
Cr$_2$BP$_3$O$_{12}$,\cite{janson2013} and apply both LSDA+$U$ and GGA+$U$
functionals.  The resulting values are given in Table~\ref{T-J}.  Both
functionals yield similar values for the interchain coupling, but the \ji\
estimates are remarkably different, especially for \lige.  Since there are no
$a~priori$ arguments favoring one of the functionals, we additionally use an independent
computational method.

Recent studies render hybrid functionals (HF) as a feasible alternative to
DFT+$U$ calculations.\cite{[{For example, }][{}]ong2011,*iori2012,*chen2012}
This method restores the insulating GS by admixing the Fock exchange into the
standard DFT exchange and correlation potential.  There is an empirical
evidence that the resulting total energies can provide accurate estimates for
the magnetic exchanges.\cite{rocquefelte2012}  Here, we use the PBE0
(Ref.~\onlinecite{pbe0}) and HSE06 (Ref.~\onlinecite{hse03,*hse06}) functionals
that are particularly suited for inorganic compounds, and involve only one free
parameter $\beta$, which controls the mixing between DFT and the Fock exchange.
To provide unbiased results, we fix this parameter to the empirically
determined optimal value $\beta$\,=\,0.25.\cite{perdew1996}  The resulting
exchange values are given in Table~\ref{T-J}.

Despite the low energy scale of magnetic couplings in both compounds, the
comparison of results obtained using different methods is quite instructive.
First, the difference between different HF schemes, PBE0 and HSE06, is
marginal, in line with previous studies.\cite{paier2006}  Second, we can
directly compare the HF results with DFT+$U$.  For \lisi, HF and GGA+$U$ yield
similar results, while LSDA+$U$ finds a substantially smaller \ji\
(Table~\ref{T-J}).  The \lige\ case is more involved.  Here, the HF estimates
are $\sim$1.5\,K smaller, than in GGA+$U$.  This difference is particularly
important for \ji: LSDA+$U$ and HF yield FM exchange, in contrast with a small
AF exchange in GGA+$U$.  Fortunately, the sign of \ji\ can be readily
determined from the experimental magnetic structure, because the
\ji--\jici--\jicii\ model is not frustrated.  Thus, the mutual arrangement of
neighboring magnetic moments in the spin chains is solely ruled by the sign of
\ji: parallel for FM \ji, and antiparallel for AF \ji.  Since the neighboring
Cr moments are antiparallel in the spin chains of both
compounds,\cite{nenert2009,nenert2010} the solutions with FM \ji\ can be ruled
out.

\begin{table}[tbp]
\caption{\label{T-J} Leading exchange integrals (in K) in \lisi\ and \lige\ as
yielded by DFT+$U$ (LSDA+$U$, GGA+$U$) and hybrid-functional (PBE0, HSE06)
calculations.  For the latter, resolving \jici\ and \jicii\ was computationally
unfeasible.  The minus sign indicates that the respective exchange coupling is FM.}
\begin{ruledtabular}
\begin{tabular}{r r r r}
\multicolumn{4}{c}{\lisi}\\ 
& \ji & \jici & \jicii \\ \hline
LSDA+$U$ (FLL, $U_d$\,=\,3\,eV) &  5.3  & 3.7 & 1.0 \\ 
GGA+$U$ (FLL, $U_d$\,=\,3\,eV) &  10.4  & 3.0 & 1.0 \\ 
PBE0                           &  11.1  & \multicolumn{2}{c}{$\Sigma = $3.7} \\
HSE06                          &  11.6  & \multicolumn{2}{c}{$\Sigma = $3.6} \\
\hline \multicolumn{4}{c}{\lige}\\ 
& \ji & \jici & \jicii \\ \hline
LSDA+$U$ (FLL, $U_d$\,=\,3\,eV) & $-4.1$ & 4.5 & 0.7 \\ 
GGA+$U$ (FLL, $U_d$\,=\,3\,eV) &   1.1  & 3.6 & 0.9 \\
PBE0                           & $-0.5$ & \multicolumn{2}{c}{$\Sigma = $2.9} \\
HSE06                          & $-0.8$ & \multicolumn{2}{c}{$\Sigma = $2.9} \\
\end{tabular}
\end{ruledtabular}
\end{table}

Regarding the interchain couplings, we note that both \jici\ and \jicii\ couple
the chains antiferromagnetically. As the pathways of these interactions involve
an additional shift along $c$ (Fig.~\ref{F-str}), the periodicities of the
crystallographic and magnetic unit cells match (the propagation vector is
$\mathbf k=0$).  On the phenomenological level, this readily yields an
effective FM interchain coupling, as indeed suggested in the earlier
studies.\cite{nenert2009,nenert2009b,nenert2010} Yet, it should be kept in mind
that this FM ``coupling'' results from two \emph{microscopic} AF exchanges
\jici\ and \jicii. 

Despite the same topology of the spin lattice, the magnetic properties of
\lisi\ and \lige\ are remarkably different.  The silicate system shows a
distinct hierarchy of the magnetic exchange couplings:
\ji\,$\gg$\,\jici\,$\gg$\,\jicii.  The dominance of the in-chain coupling \ji\
renders this compound as magnetically quasi-1D.  In contrast, a substantial
reduction of \ji\ in \lige\ leads to a physically different regime, where the
interchain coupling \jici\ overtakes the leading role.  Thus, the effective
magnetic dimensionality of \lige\ is quasi-3D.

\subsection{Model simulations}
DFT calculations provide a microscopic insight into the nature of the leading
couplings, yet the numerical accuracy does not suffice to deliver an accurate
quantitative spin model.  In the following, we adopt the DFT-based microscopic
magnetic models and refine the model parameters by simulating the
experimentally observed quantities: temperature dependence of the magnetic
susceptibility, field dependence of the magnetization, and the ordered magnetic
moment.

\begin{figure}[tb]
\includegraphics[width=8.6cm]{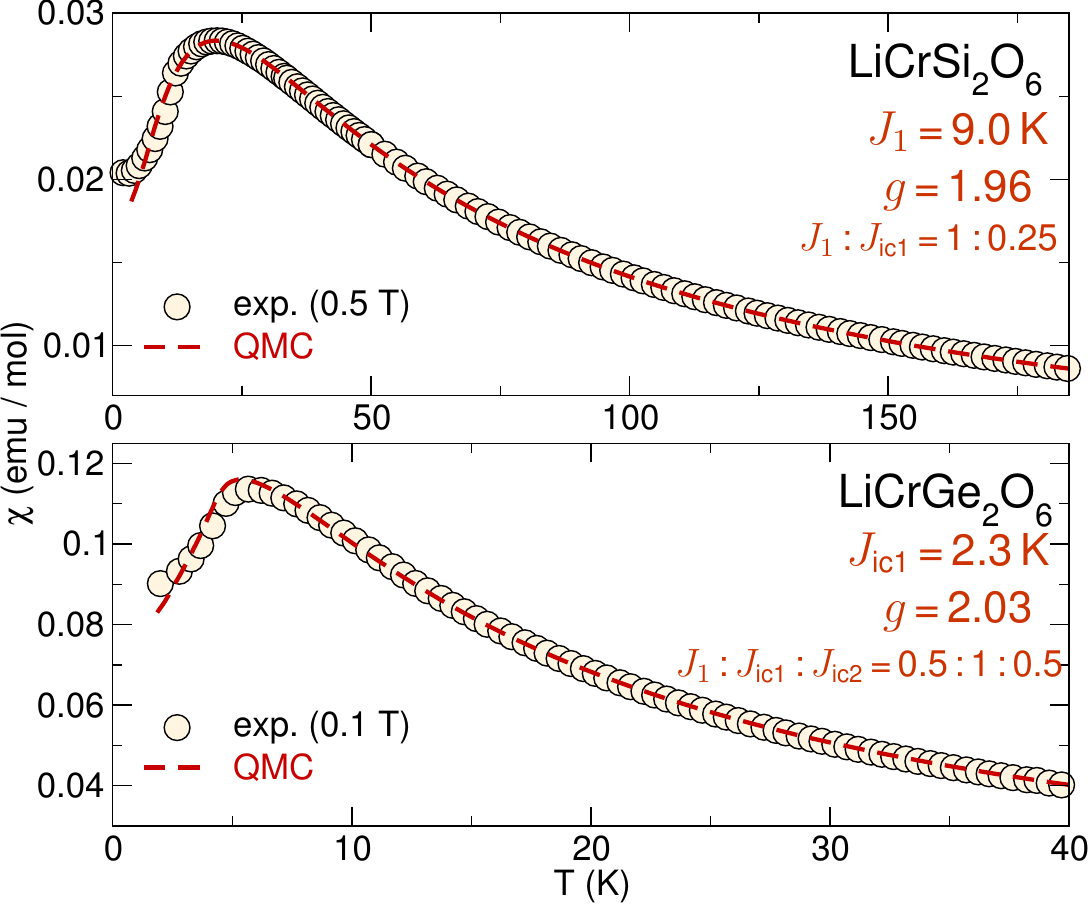}
\caption{\label{F-chi-fit} (Color
online) QMC fits (dashed line) to the experimental magnetic susceptibilities
(symbols) of \lisi\ (top) and \lige\ (bottom) measured in field of 0.5\,T and
0.1\,T, respectively.  Note that the temperature scale is different in both
panels. The leading magnetic exchange coupling amounts to \ji\,=\,9.0\,K in
\lisi\ and \jici\,=\,2.2\,K in \lige.  The ratios of the exchange couplings
amount to \ji\,:\,\jici\,=\,1\,:\,0.25 for \lisi\ and
\ji\,:\,\jici\,:\,\jicii\,=\,0.5\,:\,1\,:\,0.5 for \lige, respectively.  The
temperature-independent contribution $\chi_0$ amounts to
7.9$\times$10$^{-5}$\,emu\,/\,mol and $-4.4\times$10$^{-4}$\,emu\,/\,mol in
\lisi\ and \lige.
} \end{figure}

In \lisi, the interplane coupling \jicii\ is much smaller than \jici, thus it can
be neglected in a minimal model.  To fit the experimental magnetic
susceptibility of \lisi, we perform QMC simulations of the 2D \ji--\jici\ model,
adopting \jici:\ji\,$\simeq$\,0.25 from GGA+$U$ and HF calculations
(Table~\ref{T-J}).  We obtain an excellent fit (Fig.~\ref{F-chi-fit}) with
\ji\,=\,9.0\,K, the $g$-factor of 1.96, and the temperature-independent
contribution $\chi_0$\,=\,7.9$\times$10$^{-5}$\,emu\,/\,mol.

The magnetic model of \lige\ is very different.  First, the $|\ji|$/\jici\ ratio
is substantially smaller than in \lisi.  Second, the interchain coupling
\jicii, inactive in \lisi, is of the order of \ji\ and can not be neglected.
DFT calculations suggest that the leading coupling is \jici, yet the estimates
for \ji\ are not accurate enough to decide on the value and even on the sign of
the intrachain coupling.  To refine the values of exchange integrals, we
studied a wide range of \ji/\jici\ and \jicii/\jici\ ratios, searching for the best
possible agreement with the experimental $\chi(T)$ and the magnetic ordering
temperature.  In this way, we find that
\ji\,:\,\jici\,:\,\jicii\,=\,0.5\,:\,1\,:\,0.5 with \jici\,=\,2.3\,K and the
$g$-factor of 2.03, supplied with the temperature-independent contribution
$\chi_0$\,=\,$-4.4\times$10$^{-4}$\,emu\,/\,mol provide a good fit for the
$\chi(T)$ data above the magnetic ordering temperature (see
Fig.~\ref{F-chi-fit}), and reasonably agree with the latter (\tn\,=\,4.2\,K
versus the experimentally observed 4.8\,K).

Despite the good agreement, we can not exclude that other solutions may provide
an equally good description of the experimental data.  In general, fits to the
$\chi(T)$ data are prone to ambiguous solutions, since the magnetic
susceptibility yields information on the momentum-integrated and
thermally-averaged magnetic excitation spectrum.  For simple systems, such as
spin dimers or chains, this information suffices to evaluate the single
relevant model parameter.  For more complicated systems featuring several
relevant exchange couplings, $\chi(T)$ generally allows for ambiguous
solutions.\cite{[{For an instructive example, see }][{}] lebernegg2011}
Additional information from an independent experiment is vital to resolve this
problem.  Therefore, we perform high-field magnetization measurements that are
particularly sensitive to the structure of the magnetic excitation spectrum. The low energy scale of
magnetic couplings in \lisi\ and especially \lige\ allowed us to reach
saturation in a standard pulsed-field experiment.

\begin{figure}[tb]
\includegraphics[width=8.6cm]{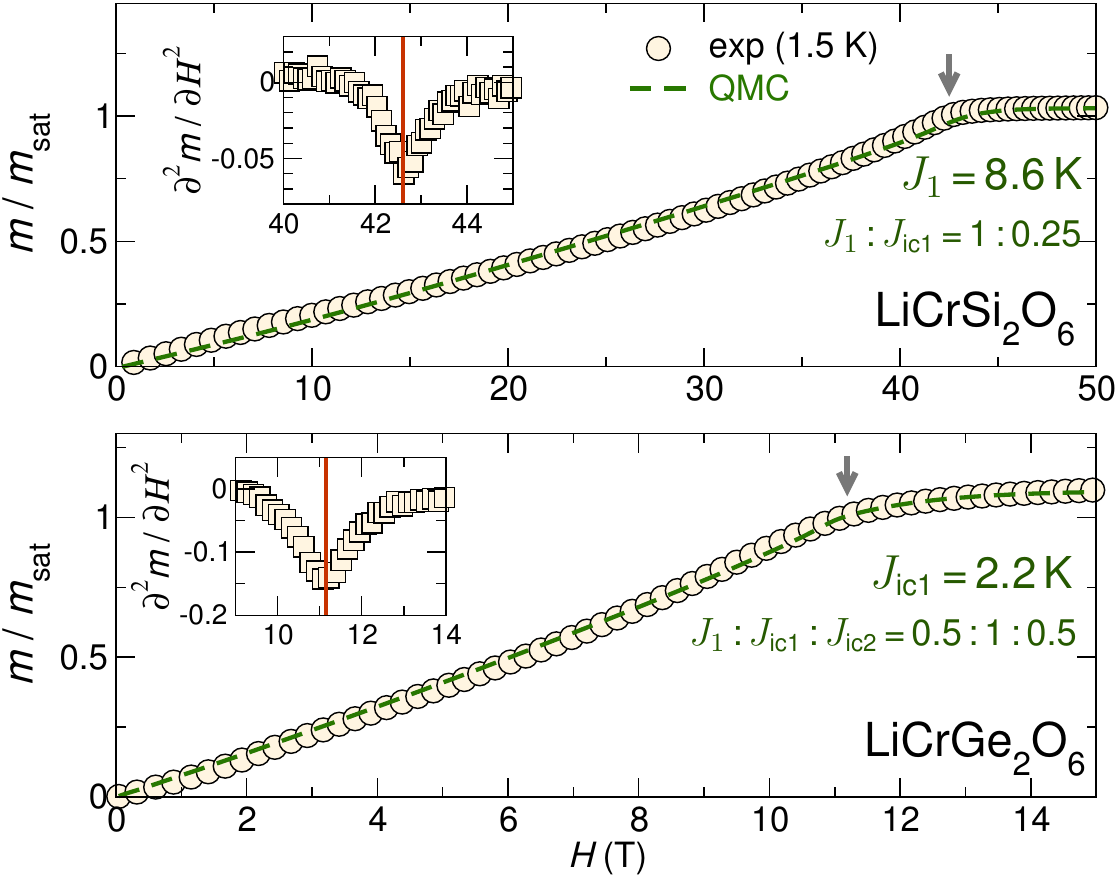}
\caption{\label{F-MH}(Color online) Magnetization curves (circles) of \lisi\
(top) and \lige\ (bottom) measured in a pulsed magnetic field at 1.5\,K.  Note
the different scales for the applied field in both panels.  Solid lines are QMC
simulations of the 2D anisotropic honeycomb lattice (\ji--\jici) model for
\lisi\ and the 3D \ji--\jici--\jicii\ model for \lige.  The ratios of the
exchange couplings amount to \ji\,\jici\,=\,1\,:\,0.25 and
\ji\,:\,\jici\,:\,\jicii\,=\,0.5\,:\,1\,:\,0.5 for \lisi\ and \lige,
respectively.  The $m_{\text{sat}}$ is the value of magnetization at the
saturation field, which is defined as a local minimum of the second derivative
(insets). 
} \end{figure}

Magnetization curves for \lisi\ and \lige\ are shown in
Fig.~\ref{F-MH}.\footnote{ The peculiar experimental setup of a pulsed-field
measurement does not allow to determine the absolute magnetization: the signal
picked up by the coil comes from a certain part of the sample, only.
Unfortunately, the ratio between the exposed and unexposed material is unknown.
Our attempts to scale the data using the low-field magnetization measurements
failed for \lisi, thus we converted the measured signal into the respective
fraction of the saturation magnetization amounting to $gS$ (Fig.~\ref{F-MH}).}
The transition to the fully polarized state (saturation) can be traced as a
local minimum in the second derivative of magnetization (Fig.~\ref{F-MH},
insets).  The different magnetic energy scales in \lisi\ and \lige\ ensue very
different saturation fields: 42.5\,T and 11.2\,T, respectively.

We simulate the $M(H)$ dependence using QMC and scale the simulated curves
using the $g$-factor values from the $\chi(T)$ fits (1.96 and 2.03 for \lisi\
and \lige, respectively).  The reduced temperature $T/\max{\{J,\jici\}}$ is
chosen to match the experimental measurement temperature, which was
$\sim$1.5\,K in both cases.  Therefore, the only adjustable parameter of the
fit is the energy scale $\max{\{\ji,\jici\}}$, which is varied in order to get
the best agreement with the experimental curve. 

In this way, we find that the \ji\,:\,\jici\,=\,1\,:\,0.25 solution  with
\ji\,=\,8.6\,K yields good agreement with the experimental magnetization
isotherm of \lisi\ (Fig.~\ref{F-MH}), justifying our restriction to the 2D
\ji--\jici\ model.  For \lige, we also obtain an excellent agreement for the
3D model with \ji\,:\,\jici\,:\,\jicii\,=\,0.5\,:\,1\,:\,0.5 and \jici\,=\,2.3\,K
(Fig.~\ref{F-MH}).  Again, QMC simulations to the $M(H)$ behavior fully
support the model assignment based on DFT calculations and $\chi(T)$ fits.

\section{\label{S-disc}Discussion}
DFT calculations reveal a substantial difference between the magnetism of
\lisi\ and \lige.  The spin model of the former features two relevant exchange
couplings, \ji\ and \jici, topologically equivalent to a honeycomb lattice
(Fig.~\ref{F-ahl}).  \lisi\ is in the quasi-1D limit of this model
(\ji\,$\gg$\,\jici), corroborated experimentally by the broad maximum in
$\chi(T)$.  In contrast, the spin model of \lige\ comprises three relevant
exchange couplings: \ji\ and \jici, as in \lisi, as well as the additional
coupling \jicii.  The ratios \ji\,:\,\jici\,:\,\jicii\ are close to
0.5\,:\,1\,:\,0.5, thus the resulting model is 3D.

\subsection{Ordered magnetic moment}
One of the main objectives of our study is to elucidate the microscopic origin
for the difference between the ordered magnetic moments on Cr atoms: 2.06\,\mb\
in \lisi\ and 2.33\,\mb\ in \lige.\cite{nenert2009,nenert2010}  For a given
spin lattice, the magnetic moment can be estimated from the respective static
structure factor $\mathbb{S}$ simulated on finite lattices.  The ordered
magnetic moment is related to $\mathbb{S}$ and the finite lattice size $N$: 

\be
m(N) = \sqrt{\frac{3\mathbb{S}}{N}}
\ee

For the finite-size scaling, we use the expression based on Eq.~(39b) of
Ref.~\onlinecite{sandvik1997}:

\be
m=\sqrt{m(N) - \frac{m_1}{\sqrt{N}} - \frac{m_2}{N}}
\ee

\begin{figure}[tbp]
\includegraphics[width=8.6cm]{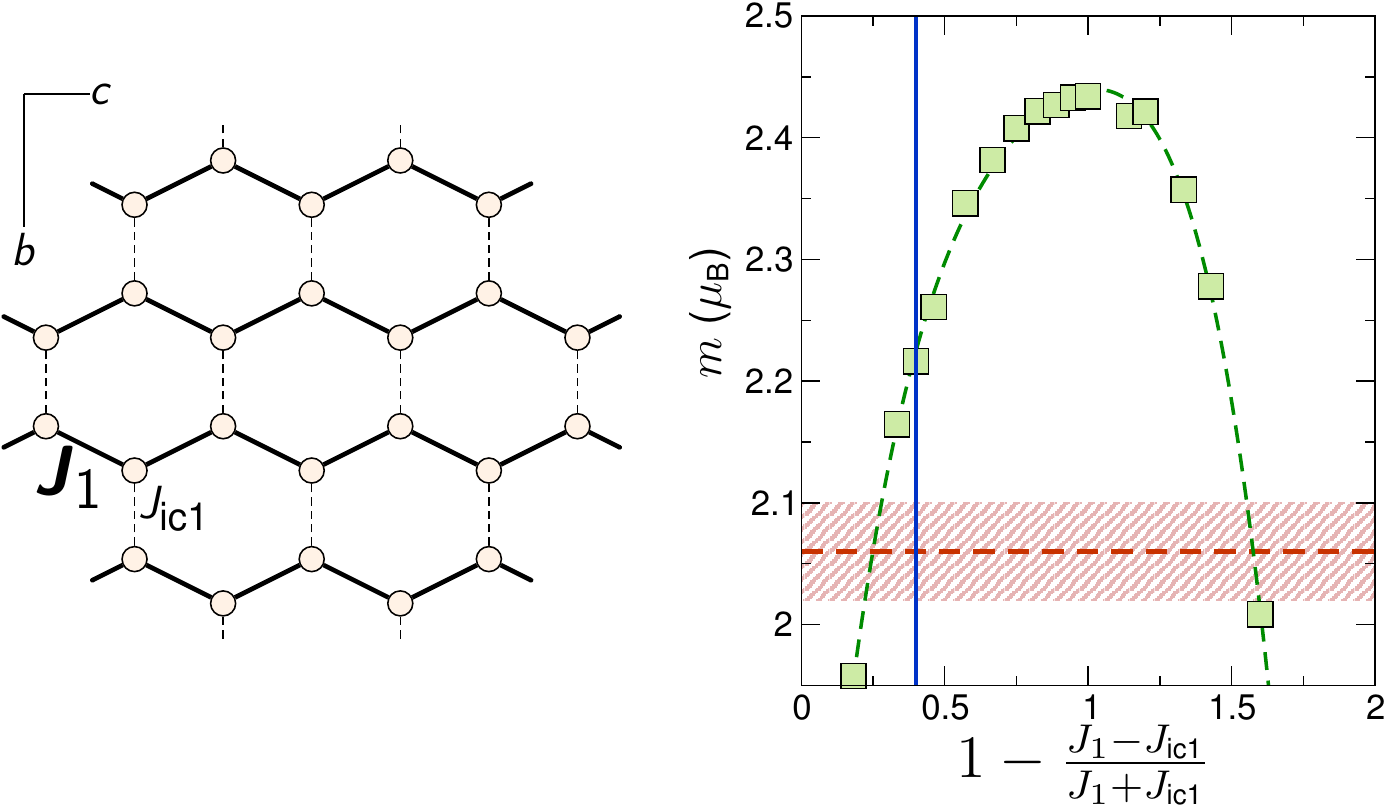}
\caption{\label{F-ahl}(Color online) Left: anisotropic honeycomb lattice model
shaped by two couplings: \ji\ along the structural chains and \jici\ between the
chains.  Right: the ordered magnetic moment $m$
in the $S$\,=\,$\frac32$ anisotropic honeycomb lattice Heisenberg model for
different $1-(\ji-\jici)/(\ji+\jici)$ ratios.  The $m$ values are evaluated as
$g\mu_{\text{B}}\langle{}S^{z}\rangle{}$ for $g$\,=\,1.96.  The vertical (blue)
line corresponds to the \ji\,:\,\jici\,=\,1\,:\,0.25 solution.  The horizontal
(red) stripe indicates the experimental magnetic moment $m_{\text{Cr}}$ in
\lisi\ determined by neutron diffraction (Ref.~\onlinecite{nenert2009}). The
dashed green curve is guide to the eye.}
\end{figure}

We first start with the 2D anisotropic honeycomb lattice model (\ji--\jici\
model, Fig.~\ref{F-ahl}), relevant for \lisi, and estimate $m$ for a broad
range of \jici\,:\,\ji\ ratios. For the $g$-factor, we adopt $g$\,=\,1.96 from
the $\chi(T)$ fits. The resulting magnetic moments are presented in the right
panel of Fig.~\ref{F-ahl}.  In the limiting case \ji\,=\,0 (\jici\,=\,0), the
magnetic model becomes 0D (1D), hence the ordered moment is exactly
zero.\cite{anderson1952} The largest moment of $m$\,=\,2.43\,\mb\ corresponds
to the isotropic case of a regular honeycomb lattice (\ji\,=\,\jici).  In the
wide range of 0.4\,$\leq$\,\jici/\ji\,$\leq$\,2.0 ratios, $m$ stays nearly
constant (the reduction does not exceed 3\,\%).  However, once the ratio
becomes smaller than 0.4 or exceeds 2.0, the ordered moment shows a sizable
reduction (Fig.~\ref{F-ahl}, right panel).

Based on our fits to $\chi(T)$ and $M(H)$ data, we concluded that the magnetism
of \lisi\ can be described by the \ji--\jici\ model with the 1\,:\,0.25 ratio of
the leading couplings, depicted as the vertical line in the right panel of
Fig.~\ref{F-ahl}.  The respective simulated magnetic moment of $\sim$2.23\,\mb\
is significantly larger than the experimentally determined \mcr\,=\,2.06(4)\,\mb.
Similarly, the ordered magnetic moment can be evaluated for the
\ji--\jici--\jicii\ model realized in \lige.  In this way, adopting
\ji\,:\,\jici\,:\,\jicii\,=\,0.5\,:\,1\,:\,0.5 and $g$\,=\,2.03, we obtain
$m$\,$\simeq$\,2.63\,\mb, again substantially larger than
\mcr\,=\,2.33(3)\,\mb\ determined experimentally. 

Despite the overestimation of the absolute values, the difference between $m$
for \lisi\ and \lige\ is about 0.4\,\mb, which reasonably agrees with the
$\sim$0.25\,\mb\ difference found experimentally.  Thus, different \mcr\ in
\lisi\ and \lige\ originate from the different dimensionality of the underlying
microscopic model, and not from magnetic frustration, as speculated
earlier.

The overestimation of magnetic moments in QMC can be assigned from the partial
moment transfer from Cr to the surrounding O atoms, which is neglected in the
simulation. This transfer, commonly referred to as the covalency effect, was
experimentally demonstrated to reduce $m_{\text{Cr}}$ in Cr$_2$O$_3$.\cite{brown1993}
The reported reduction of $\sim$0.25\,\mb\ is close to the difference between
the experimental values of $m_{\text{Cr}}$ in \lix\ and the respective QMC
estimates.

\begin{figure}[tb]
\includegraphics[width=8.6cm]{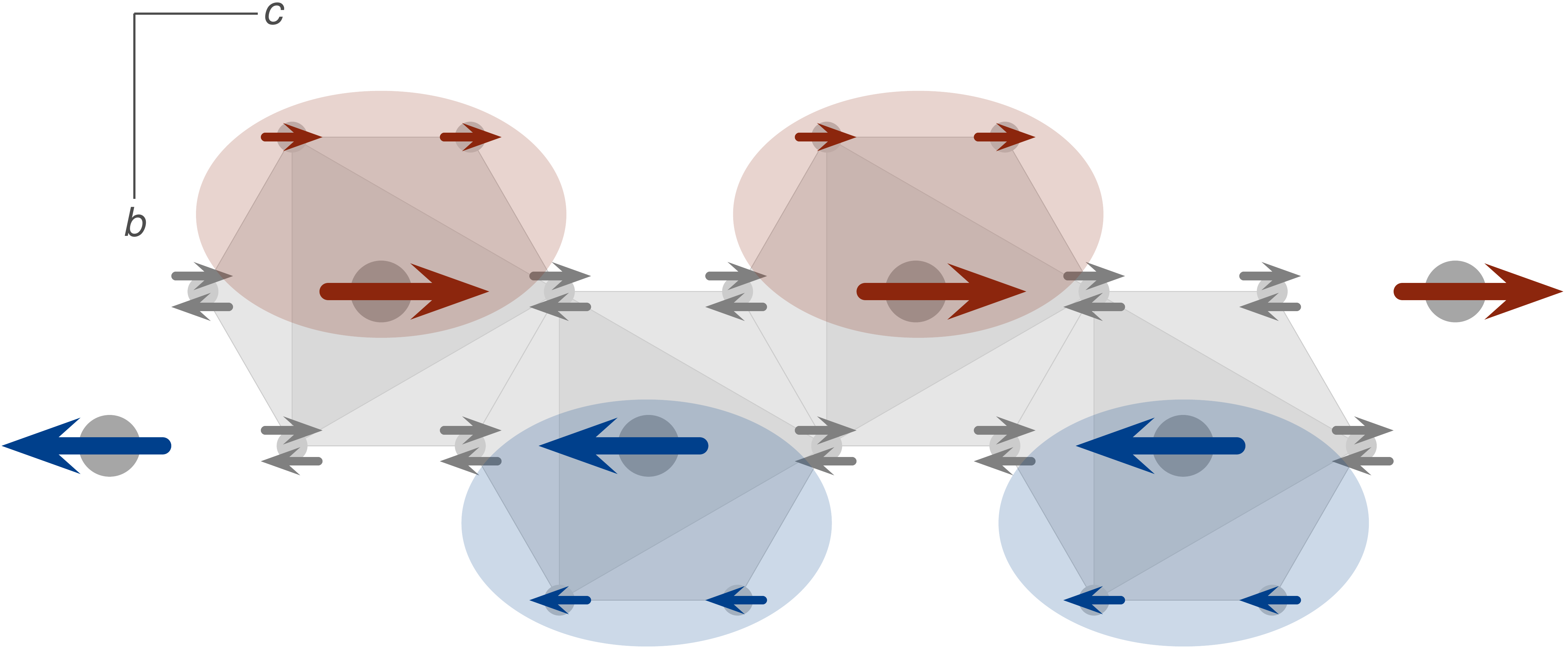}
\caption{\label{F-moment-reduc}(Color online) Magnetic chains of edge-sharing
CrO$_6$ octahedra in the AF ordered state.  Magnetic moments
on Cr (O) atoms are depicted with large (small) arrows.  For each Cr, only two
neighboring O atoms carry non-zero magnetic moment.  As a result, the local
magnetization density is shifted along the $b$ axis.
} \end{figure}

The peculiar feature of pyroxenes is the inequivalence of six oxygens,
surrounding the magnetic Cr atom:  Four out of these six O atoms are shared by
neighboring Cr atoms, hence the induced moments cancel out in
the AF state (Fig.~\ref{F-moment-reduc}).  The remaining two O atoms carry
a non-zero moment.  For instance, our GGA+$U$ calculations with $U_d=3$\,eV yield small moments
of $m_{\text{O}}$\,$\simeq$\,0.02\,\mb, which is twice smaller than
the value required to account for the $\sim$0.25\,\mb\ reduction of
$m_{\text{Cr}}$.  Due to the limited sensitivity of computational DFT-based
schemes to low magnetic moments, we propose a direct experimental measurement
of $m_{\text{O}}$.  Although such small moments are beyond the resolution of
standard neutron scattering techniques, they can be estimated by neutron
polarimetry.\cite{brown1993}

Non-zero $m_{\text{O}}$ can have significant ramifications, which we did not
consider so far.  As can be seen in Fig.~\ref{F-moment-reduc}, the concomitant
effect of the magnetic moment transfer from Cr to O is the shift of
magnetization density along the $b$ axis.  Interestingly, this coincides with
the direction of the maximal electric polarization.\cite{jodlauk2007}
Therefore, experimental measurement of $m_{\text{O}}$ can be important for
understanding the ME effect in the \lixx\ compounds.

\subsection{Magnetostructural correlations}
The combination of the small magnetic energy scale and the rich crystal chemistry
of pyroxenes is advantageous for tuning the magnetic properties.  However, a
directed tuning is possible only if the relevant structural parameters are
identified and the respective magnetostructural correlations are known.  To
gain better understanding of these correlations, we evaluate the mechanisms
underlying intrachain as well as interchain couplings in Cr pyroxenes.

The nearest-neighbor exchange $J_1$ operates between two CrO$_6$ octahedra
sharing a common O..O edge.  For the AF part of the exchange, the leading
hopping is mediated by the two $t_{2g}$ orbitals lying in the same plane with
the common O..O edge: $xz$ on one Cr and $yz$ on the other.
\footnote{Different orbital characters ensue from the peculiar symmetry of
\lix.  Here, we adopt local coordinate frame with the $x$ and $y$ axes running
towards the O2a and O2b atoms. This leads to different frames on the
neighboring Cr atoms:the $x$ axis on one Cr transforms into the $-y$ axis on
the neighboring Cr, the $y$ transforms into the $-x$, while $z$ changes its
sign.} Since these two orbitals are $\sigma$-overlapping, they facilitate a
direct $d$--$d$ hopping, thus the Cr--Cr distance $d_{\text{(Cr--Cr)}}$ is the
key parameter, determining the strength of magnetic exchange, as was already
pointed out in Ref.~\onlinecite{streltsov2008}.  It should be noted, however,
that the FM contribution to $J_1$ also depends on the Cr--Cr distance.  For
instance, the leading $t_{2g}\!\leftrightarrow\!e_g$ hopping in \lige\ is
reduced by about 20\% compared to the respective term in \lisi.  This is
smaller than $\sim$27\% difference in the leading $t_1^{xz\leftrightarrow{}yz}$
terms, responsible for the AF exchange, hence the total exchange $J_1$ reduces.
This perfectly agrees with larger $J_1$ in \lisi\
($d_{\text{(Cr--Cr)}}$\,$\simeq$\,3.052\,\r{A}) than in \lige\
($d_{\text{(Cr--Cr)}}$\,$\simeq$\,3.101\,\r{A}), and could explain the FM
exchange in NaCrGe$_2$O$_6$ ($d_{\text{(Cr--Cr)}}$\,$\simeq$\,3.140\,\r{A}).

We turn now to the magnetic exchange between the chains.  In contrast to
earlier conjectures, our microscopic analysis discloses that the interchain
coupling is realized by two paths involving double bridges of anionic XO$_4$
tetrahedra (\jici\ and \jicii). Other interchain paths, including single
bridges of the XO$_4$ tetrahedra, are essentially inactive. 

The orbitals of Si and Ge provide minor contributions to the Cr $3d$ bands
(Fig.~\ref{F-dos}), thus the \mbox{Cr--O..O--Cr} interaction pathways are
expected to be most relevant. The nature of interacting orbitals can be
understood from the crystal structure. The coupling \jici\ requires one of the
$t_{2g}$ orbitals ($yz$) that overlaps with the $p$ orbitals of O1b and O2b.
Likewise, \jicii\ involves the hopping processes through O1a and O2a, so that
the $xz$ orbital becomes active.\cite{suppl} Altogether, each of the \jici\ and
\jicii\ exchanges entails only one out of three half-filled orbitals of
Cr$^{3+}$. 

\begin{figure}[tb]
\includegraphics[width=8.6cm]{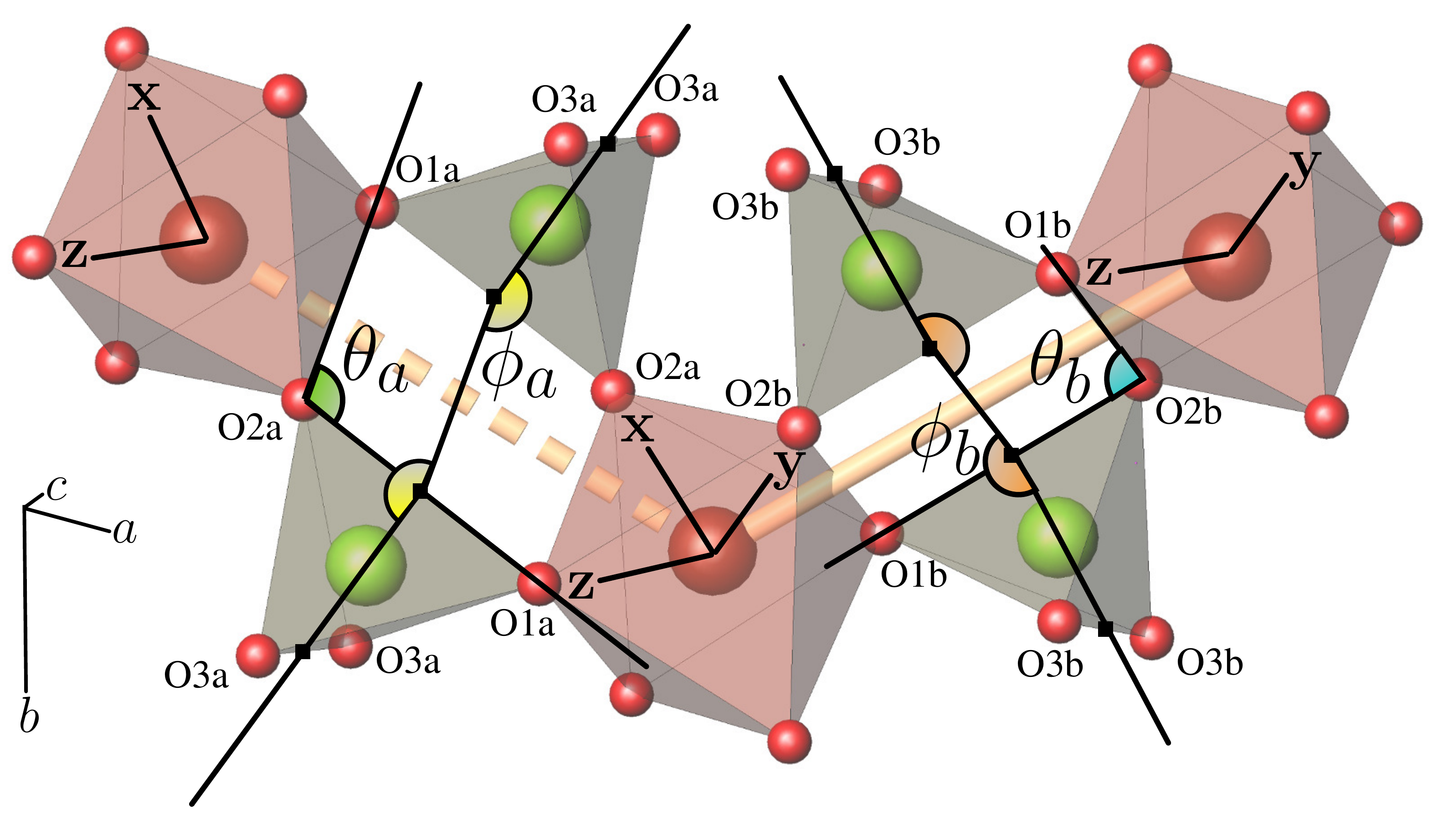}
\caption{\label{F-interchain}(Color online) Superexchange paths facilitating
the leading interchain couplings \jici\ (thick solid cylinder) and \jicii\
(thick sectioned cylinder) in \lix.  Similar to Fig.~\ref{F-str}, both paths
are mediated by double bridges of anionic $X$O$_4$ tetrahedra.  The labels $\mathbf{x}$,
$\mathbf{y}$, and $\mathbf{z}$ denote the local coordinate axes.
$\phi_a$ and $\phi_b$ angles measure the rotation of anionic tetrahedra with
respect to the Cr--Cr axis.  $\theta_a$ and $\theta_b$ measure the lateral
shift of CrO$_4$ planes of the neighboring CrO$_6$ octahedra.  Numerical values
for $\phi_a$, $\phi_b$,  $\theta_a$, and  $\theta_b$ are provided in
Table~\ref{T-interchain}.}
\end{figure}

At this point, an unexpected connection to V$^{4+}$ phosphates can be drawn,
because V$^{4+}$ features only one half-filled (magnetic) orbital. It is widely
accepted that phosphorous states do not contribute to the superexchange in
phosphates,\cite{roca1998,janson2011} hence the V--O..O--V interaction pathways
are envisaged. Similar to pyroxenes, double bridges of the PO$_4$ tetrahedra
are more efficient than single bridges.\cite{petit2003,tsirlin2011} According
to Ref.~\onlinecite{roca1998}, several geometrical parameters are of crucial
importance for the superexchange through these double bridges: (i) the in-plane
shift of the MO$_6$ octahedra (shear deformation of the O$_4$ square shared by
these polyhedra); (ii) the vertical shift of the MO$_6$ octahedra (normal
deformation of O$_4$); and (iii) the rotation of PO$_4$ tetrahedra around the
O..O edges of O$_4$, parallel to the M--M line (here, we use M to denote the
magnetic ion, either V$^{4+}$ in Ref.~\onlinecite{roca1998} or Cr$^{3+}$ in our
case). In pyroxenes, the vertical shift is nearly absent, while the in-plane
shift and the rotations of the tetrahedra can be quantified by the angles
$\theta$ and $\phi$, respectively.\footnote{The angle $\phi_\gamma$ is measured
between the line connecting the midpoints of two O2$\gamma$..O1$\gamma$ edges
and the midpoint of O2$\gamma$..O1$\gamma$ edge with that of
O3$\gamma$..O3$\gamma$ edge (Fig.~\ref{F-interchain}), where $\gamma$\,=\,a or
b} The largest AF superexchange is expected at $\theta=90^{\circ}$ and
$\phi=180^{\circ}$. Any deviations from this fully symmetric configuration
reduce the AF exchange, with the in-plane shift ($\theta>90^{\circ}$) having
the most pronounced effect on the coupling.\cite{roca1998} 

The relevant geometrical parameters for \jici\ and \jicii\ are summarized in
Table~\ref{T-interchain}. The difference between \jici\ and \jicii\ is well in
line with the $\theta_{\gamma}$ values: the larger in-plane shift for \jicii\
reduces this coupling compared to \jici. The orientation of the tetrahedra
($\phi_{\gamma}$) implies an opposite trend, yet this geometrical parameter
seems to be of relatively low importance for the superexchange in pyroxenes,
similar to the V$^{4+}$ phosphates discussed in Ref.~\onlinecite{roca1998}. It
is worth noting that the $d_{\text{O..O}}$ distance, the edge of the XO$_4$
tetrahedron, has no direct influence on the magnetic coupling. Both \lisi\ and
\lige\ reveal very similar interchain interactions, even though the GeO$_4$
tetrahedra are much bigger than their SiO$_4$ counterparts, and the interacting
oxygen atoms (O1$\gamma$ and O2$\gamma$) are further apart. This unexpected
behavior shows that the Cr--O..O--Cr superexchange does not involve individual
oxygen orbitals. It rather pertains to an interaction between the Cr $d$
orbital and one of the molecular orbitals of the XO$_4$ tetrahedron. The
interaction between O1$\gamma$ and O2$\gamma$ is largely determined by
electronic interactions within the XO$_4$ tetrahedron, whereas the
superexchange is a more subtle effect that relies on the mutual orientations of
the CrO$_6$ octahedra and XO$_4$ tetrahedra.

We can also draw more general conclusions regarding microscopic aspects of
magnetic pyroxenes. The replacement of Si with Ge changes the Cr--Cr distance
within the chains and, consequently, the intrachain coupling $J_1$, whereas the
regime of the interchain couplings with \jici$>$\jicii\ is largely retained.
The Li/Na substitution should have a stronger effect on both intrachain and
interchain couplings. The size difference between Li$^{+}$ and Na$^{+}$ will
further alter the Cr--Cr distance and eventually render $J_1$
FM.\cite{streltsov2008,nenert2009b} Additionally, Na-based pyroxenes feature a
higher crystallographic symmetry ($C2/c$) that makes \jici\ and \jicii\
equivalent. 

\begin{table}[tb]
\caption{\label{T-interchain} Magnetostructural correlations for the interchain
exchanges \jicii\ and \jici.  For the corresponding structural elements, see
Fig.~\ref{F-interchain}. $d_{\text{O..O}}$ is the length (in \r{A}) of
O1$\gamma$..O2$\gamma$ edge of the XO$_4$ tetrahedron. $\phi_\gamma$ is the
angle (in $^{\circ}$) between the line connecting the midpoints of two
O2$\gamma$..O1$\gamma$ edges and the midpoint of O2$\gamma$..O1$\gamma$ edge
with that of O3$\gamma$..O3$\gamma$ edge.  $\theta_\gamma$ is the
O1$\gamma$--O2$\gamma$--O1$\gamma$ angle (in $^{\circ}$).  Numerical values for
the exchange couplings are based on QMC simulations and GGA+$U$ calculations.}
\begin{ruledtabular}
\begin{tabular}{l c r r c r r}
\multicolumn{1}{c}{compound} & \multicolumn{2}{c}{exchange} &
\multicolumn{1}{c}{$d_{\text{O..O}}$} &  \multicolumn{1}{c}{$\gamma$} &
\multicolumn{1}{c}{$\phi_\gamma$} & \multicolumn{1}{c}{$\theta_\gamma$}\\
\hline \multirow{2}{*}{\lisi}  & \jici    & $\sim$2.2\,K      & 2.7574 & a   &
102.3 &  98.9\\
                        & \jicii & $\lesssim$1.0\,K  & 2.7171 & b   & 114.3 & 109.2\\
\multirow{2}{*}{\lige}  & \jici    & $\sim$2.3\,K      & 2.9801 & a   &  94.8 &  93.3\\
                        & \jicii & $\sim$1.1\,K      & 2.9299 & b   & 114.2 & 113.2\\
\end{tabular}
\end{ruledtabular}
\end{table}

\section{\label{S-summary}Summary}

Using a combination of density functional theory (DFT) calculations and quantum
Monte Carlo (QMC) simulations, we evaluate the microscopic magnetic models for
two $S$\,=\,3/2 pyroxenes, \lisi\ and \lige.  The magnetism of \lisi\ is
characterized by a 2D anisotropic honeycomb lattice model, with
\ji\,$\simeq$\,8\,K running along the structural chains and \jici\ between the
chains.  Since \ji\,$\gg$\,\jici, \lisi\ is very close to the 1D limit of this
model.  In contrast, the spin model of \lige\ is 3D, with
\ji\,$\simeq$\,1.1\,K, \jici\,$\simeq$\,2.3\,K, and an additional interplane
coupling \jicii\,$\simeq$\,1.1\,K.  Here, the strongest magnetic exchange \jici\
operates between the structural chains.  Both spin models lack magnetic
frustration, in contrast with earlier speculations.

Despite very low energies of the individual magnetic couplings in Cr$^{3+}$-based
pyroxenes, we were able to obtain quantitative microscopic magnetic models.
This would not be possible without a combination of DFT calculations and
experiment, because the DFT results alone remain rather ambiguous and sensitive
to the choice of the computational parameters, which are connected with the
approximate treatment of the strong correlations in the $3d$ shell. The
modeling of magnetic susceptibility together with the high-field magnetization
data helps us to refine the model parameters and eventually explain other
observable properties, such as the different ordered magnetic moment in \lisi\
and \lige. We believe that a similar approach can be applied to other magnetic
pyroxenes, including the Fe$^{3+}$-based multiferroic compounds, and we hope
that our work will
stimulate further studies in this direction.

By comparing \lisi\ and \lige, we conclude that the interchain couplings in
Cr$^{3+}$-based pyroxenes are typically AF, non-frustrated, and run via double
bridges of the XO$_4$ tetrahedra. This is very different from the earlier
assumptions of ferromagnetic interchain interactions, even though the
interchain order is indeed ferromagnetic in the sense that it matches the
periodicity of the crystal structure. The overall crystallographic symmetry is
also very important for the magnetism of pyroxenes.

The Li compounds with the lower $P2_1/c$ symmetry feature inequivalent
interchain couplings and may even show an unexpected quasi-2D magnetism when
\jici$\gg$\jicii. By contrast, the higher $C2/c$ symmetry of the Na compounds
should render the magnetic system 3D with possible 1D features at
$J_1\!\gg\!J_{\text{ic1}}\!\equiv\!J_{\text{ic2}}$.

\acknowledgments
Fruitful discussions with D. I. Khomskii are gratefully acknowledged.  OJ was
supported by the Mobilitas program of the ESF, grant number MJD447, and the
PUT210 grant of the Estonian Research Council.  AT was supported via the ESF
Mobilitas program, grant number MTT77. Discussions with Martin Rotter are
kindly acknowledged. We also acknowledge the support of the HLD at HZDR, member
of the European Magnetic Field Laboratory (EMFL). 

%

\setcounter{figure}{0}
\setcounter{table}{0}
\renewcommand{\thefigure}{S\arabic{figure}}
\renewcommand{\thetable}{S\arabic{table}}

\begin{widetext}
\newpage
\begin{center}
\thispagestyle{empty}
{\large
Supplementary information for 
\smallskip

\textbf{Magnetic pyroxenes LiCrGe$_2$O$_6$ and LiCrSi$_2$O$_6$:\\ dimensionality
crossover in a non-frustrated $S$\,=\,$\frac{3}{2}$ Heisenberg model}
\medskip

\normalsize O. Janson, G. N\'enert, M. Isobe, Y. Skourski, Y. Ueda, H. Rosner, and A. A. Tsirlin}
\end{center}
\medskip

\vskip -1cm 

\begin{table}[h]
\caption{\label{T_WF} Leading transfer integrals $t_i^{mm'}$ ($m$ and $m'$ are
orbital indices) in \lisi\ and \lige, evaluated by mapping the GGA band
structure onto the Wannier basis.  All values are given in~meV.  The coordinate
systems are local (Cr-centered), with the $x$ and $y$ axes running towards the
neighboring O2a and O2b atoms, respectively (see Fig.~\ref{F-WF}, (a)). The
leading terms are shown bold.  Color denotes whether the respective term
contributes to {\cred antiferromagnetic} or {\cblue ferromagnetic} exchange.
Note that for $t_1$, the hopping matrix is not symmetric
($t_1^{mm'}\!\neq\!t_1^{m'm}$) due to the absence of centre of inversion
between the respective Cr atoms.  In contrast, both interchain paths pass
through the center of inversion, thus
$t_{ij}^{mm'}\!=\!\pm{}t_{ij}^{m'm}$.}
\begin{ruledtabular}
\begin{tabular}{r r r r r r r r r r r r}
& \multicolumn{5}{c}{\lisi} & & \multicolumn{5}{c}{\lige}\\ \hline
& \multicolumn{5}{c}{$t_1$ ($d_{\text{Cr--Cr}}$\,=\,3.052\,\r{A})} & &
\multicolumn{5}{c}{$t_1$ ($d_{\text{Cr--Cr}}$\,=\,3.101\,\r{A})}\\
& $|xy\rangle$ & $|xz\rangle$ & $|yz\rangle$ & $|z^2-r^2\rangle$ & $|x^2-y^2\rangle$ & 
& $|xy\rangle$ & $|xz\rangle$ & $|yz\rangle$ & $|z^2-r^2\rangle$ & $|x^2-y^2\rangle$ \\
$\langle{}xy|$     &  $-50$ & $-18$ & $54$ & -- & $-12$ &
$\langle{}xy|$     &  $-37$ & $-26$ & $66$ & $-18$ & -- \\
$\langle{}xz|$     &  $-60$ & $-15$ & $53$ & -- & $-15$ &
$\langle{}xz|$     &  $-59$ & $-25$ & $45$ & $12$ & $-23$ \\
$\langle{}yz|$     &  $19$ & {\cred $\mathbf{135}$} & -- & $-90$ & {\cblue $\mathbf{-140}$} &
$\langle{}yz|$     &  $13$ & {\cred $\mathbf{98}$} & $-16$ & $-108$ & {\cblue $\mathbf{-115}$} \\
$\langle{}z^2-r^2|$&  $32$ & $100$ & -- & $24$ & $-59$ &
$\langle{}z^2-r^2|$&  $62$ & $73$ & $10$ & $69$ & $-89$ \\
$\langle{}x^2-y^2|$&  -- & {\cblue $\mathbf{-116}$} & $12$ & $43$ & -- &
$\langle{}x^2-y^2|$&  -- & {\cblue $\mathbf{-108}$} & $31$ & $83$ & $10$  \\ \hline
& \multicolumn{5}{c}{$t_{\text{ic1}}$ ($d_{\text{Cr--Cr}}$\,=\,5.336\,\r{A})} & &
\multicolumn{5}{c}{$t_{\text{ic1}}$ ($d_{\text{Cr--Cr}}$\,=\,5.581\,\r{A})}\\
& $|xy\rangle$ & $|xz\rangle$ & $|yz\rangle$ & $|z^2-r^2\rangle$ & $|x^2-y^2\rangle$ &
& $|xy\rangle$ & $|xz\rangle$ & $|yz\rangle$ & $|z^2-r^2\rangle$ & $|x^2-y^2\rangle$ \\
$\langle{}xy|$     &  -- & $11$ & $-12$ & -- & -- &
$\langle{}xy|$     &  -- & -- & -- & $31$ & -- \\
$\langle{}xz|$     &  $-11$ & -- & -- & -- & $-12$ &
$\langle{}xz|$     &  -- & -- & $-10$ & -- & -- \\
$\langle{}yz|$     &  $-12$ & -- & {\cred $\mathbf{62}$} & $-26$ & {\cblue $\mathbf{-53}$} &
$\langle{}yz|$     &  -- & $10$ & {\cred $\mathbf{67}$} & $-17$ & {\cblue $\mathbf{-66}$} \\
$\langle{}z^2-r^2|$&  -- & -- & $26$ & $68$ & $59$ &
$\langle{}z^2-r^2|$&  $-31$ & -- & $17$ & $121$ & $85$ \\
$\langle{}x^2-y^2|$&  -- & $-12$ & {\cblue $\mathbf{53}$} & $59$ & -- &
$\langle{}x^2-y^2|$&  -- & -- & {\cblue $\mathbf{66}$} & $85$ & $-14$ \\ \hline
& \multicolumn{5}{c}{$t_{\text{ic2}}$ ($d_{\text{Cr--Cr}}$\,=\,5.322\,\r{A})} & &
\multicolumn{5}{c}{$t_{\text{ic2}}$ ($d_{\text{Cr--Cr}}$\,=\,5.464\,\r{A})}\\
& $|xy\rangle$ & $|xz\rangle$ & $|yz\rangle$ & $|z^2-r^2\rangle$ & $|x^2-y^2\rangle$ & 
& $|xy\rangle$ & $|xz\rangle$ & $|yz\rangle$ & $|z^2-r^2\rangle$ & $|x^2-y^2\rangle$ \\
$\langle{}xy|$     &  -- & $-13$ & $19$ & $-18$ & -- & 
$\langle{}xy|$     &  -- & -- & $11$ & $-13$ & -- \\
$\langle{}xz|$     &  $13$ & {\cred $\mathbf{-52}$} & -- & $32$ & {\cblue $\mathbf{-42}$} &
$\langle{}xz|$     &  -- & {\cred $\mathbf{-66}$} & -- & $47$ & {\cblue $\mathbf{-78}$} \\
$\langle{}yz|$     &  $19$ & -- & -- & -- & $-10$ &
$\langle{}yz|$     &  $11$ & -- & -- & -- & -- \\
$\langle{}z^2-r^2|$&  $18$ & $32$ & -- & $44$ & $-40$ &
$\langle{}z^2-r^2|$&  $13$ & $47$ & -- & $74$ & $-66$ \\
$\langle{}x^2-y^2|$&  -- & {\cblue $\mathbf{-42}$} & $10$ & $-40$ & -- &
$\langle{}x^2-y^2|$&  -- & {\cblue $\mathbf{-78}$} & -- & $-66$ & -- \\
\end{tabular}
\end{ruledtabular}
\end{table}

\begin{figure}[h]
\includegraphics[width=.67\textwidth]{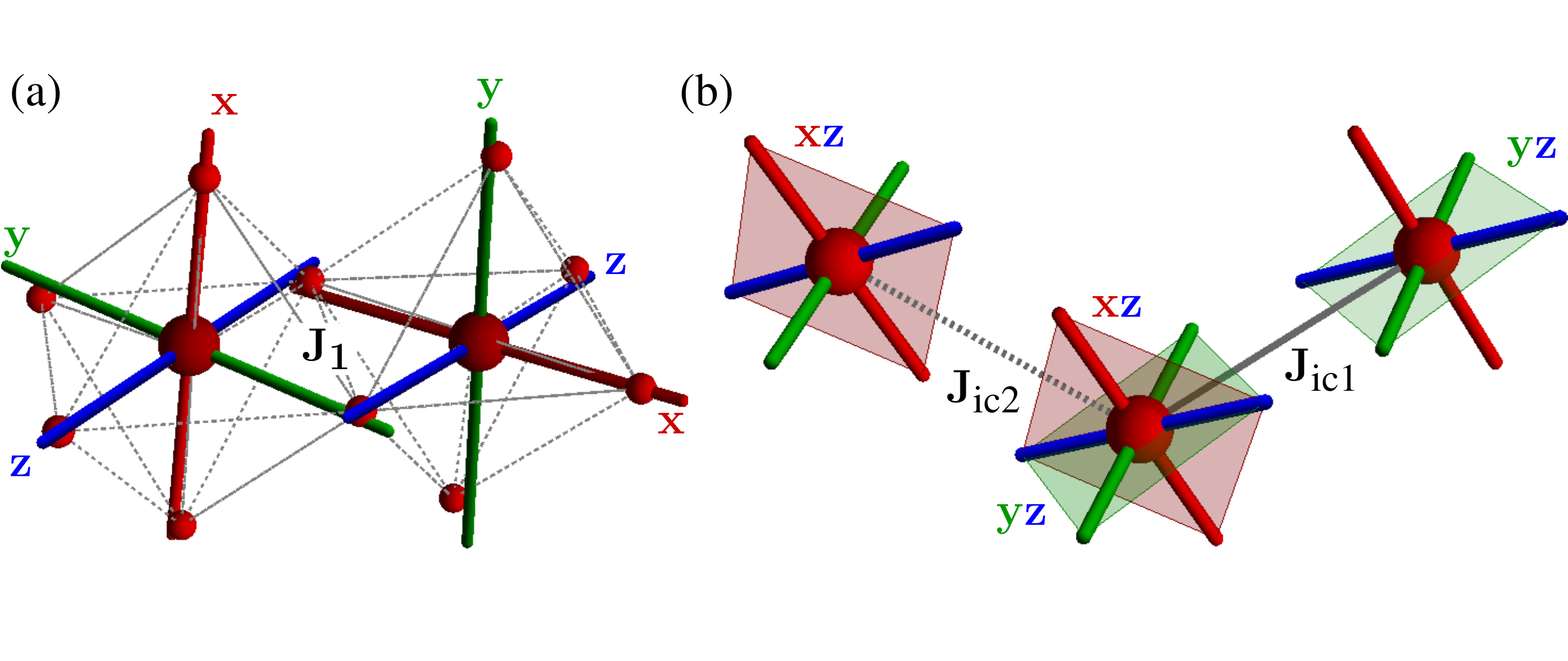}
\caption{\label{F-WF} Local coordinate systems of CrO$_6$ tetrahedra used for
construction of the Wannier functions. (a) The antiferromagnetic part of
nearest-neighbor exchange $J_1$ is dominated by the direct overlap of $xz$ and
$yz$ orbitals. (b) The antiferromagnetic exchange between the chains is ruled
by the superexchange between nearly in-plane $yz$($xz$) orbitals for the
$J_{\text{ic1}}$($J_{\text{ic2}}$) coupling.
}
\end{figure}

\end{widetext}

\end{document}